%% file: Group32_04.tex
\documentclass[a4paper]{jpconf}
\usepackage[utf8]{inputenc}
\usepackage{iopams}
\usepackage{graphicx}
\usepackage{amsmath}
\usepackage{times}
\usepackage{tensor}
\usepackage{float}
\newcommand{\RB}{\mathbb{R}}
\newcommand{\Hf}{H}
\newcommand{\HCd}{\mathcal{\Hf}}
\newcommand{\onehalf}{{\textstyle\frac{1}{2}}}
\newcommand{\twothird}{{\textstyle\frac{2}{3}}}
\newcommand{\onethird}{{\textstyle\frac{1}{3}}}
\newcommand{\quarter}{{\textstyle\frac{1}{4}}}
\newcommand{\threequarter}{{\textstyle\frac{3}{4}}}
\newcommand{\afffias}{Frankfurt Institute for Advanced Studies (FIAS), Ruth-Moufang-Strasse~1, 60438 Frankfurt am Main, Germany}
\newcommand{\affjwg}{Goethe-Universit\"at, Max-von-Laue-Strasse~1, 60438~Frankfurt am Main, Germany}
\newcommand{\affgsi}{GSI Helmholtzzentrum f\"ur Schwerionenforschung GmbH, Planckstrasse~1, 64291 Darmstadt, Germany}
\begin{document}
\title{Covariant Canonical Gauge Gravitation and Cosmology}
\author{D Vasak$^1$, J Kirsch$^1$, D Kehm$^{1,2}$ and J Struckmeier$^{1,2,3}$}
\address{$^1$ \afffias}
\address{$^2$ \affjwg}
\address{$^3$ \affgsi}
\ead{vasak@fias.uni-frankfurt.de}

\begin{abstract}%
The covariant canonical transformation theory applied to the relativistic Hamiltonian theory of classical matter fields in dynamical space-time yields a novel (first order) gauge field theory of gravitation. The emerging field equations necessarily embrace a quadratic Riemann term added to Einstein's linear equation. The quadratic term endows space-time with inertia generating a dynamic response of the space-time geometry to deformations relative to (Anti) de~Sitter geometry. A ``deformation parameter'' is identified, the inverse dimensionless coupling constant governing the relative strength of the quadratic invariant in the Hamiltonian, and directly observable via the deceleration parameter $q_0$. The quadratic invariant makes the system inconsistent with Einstein's constant cosmological term, $\Lambda = \mathrm{const}$. In the Friedman model this inconsistency is resolved with the scaling ansatz of a ``cosmological function'', $\Lambda(a)$, where $a$ is the scale parameter of the FLRW metric.
The cosmological function can be normalized such that with the $\Lambda$ CDM parameter set the present-day observables, the Hubble constant and the deceleration parameter, can be reproduced.
With this parameter set we recover the dark energy scenario in the late epoch. The proof that inflation in the early phase is caused by the ``geometrical fluid'' representing the inertia of space-time is yet pending, though.
Nevertheless, as according to the CCGG theory the present-day cosmological function, identified with the currently observed $\Lambda_\mathrm{obs}$, is a balanced mix of two contributions. These are the (A)dS curvature plus the residual vacuum energy of space-time and matter. The curvature term is proportional to the deformation parameter given by the coupling strength of the quadratic Riemann term. This allows for a fresh look at the Cosmological Constant Problem that plagues the standard Einstein-Friedman cosmology.

\medskip
Presented at the 32nd International Colloquium on Group Theoretical Methods in Physics (Group32) in Prague, Czech Republic, in July 2018.
\end{abstract}




\section{Introduction} \label{sec1}
Covariant canonical transformation theory is a rigorous framework to enforce symmetries of classical relativistic fields with respect to arbitrary transformation (Lie) groups~\cite{struckmeier08, struckmeier13}.  The imposed requirement of invariance of the original action integral with respect to local symmetry transformations is ``cured'' via the introduction of additional degrees of freedom, the gauge fields. This cure closes the system with the gauge fields emerging as the mediators of forces acting upon the original fields (or particles). Those forces can be regarded as pseudo forces, like the centrifugal force ``felt'' by particles moving on a circle. However, here these force fields acquire the status of dynamic entities and, in the language of field theory, of independent fundamental fields.

The canonical transformation theory applied to matter fields embedded in dynamic space-time  leads to a novel gauge theory of gravitation (Covariant Canonical Gauge theory of Gravitation, CCGG) that contains also the linear Einstein-Hilbert (EH) term~\cite{struckvasak15, struckmeier17a, struckmeier18a}. It implements the postulate of General Relativity (GR), requesting invariance of the action integral with respect to arbitrary transition maps of space-time co-ordinates (diffeomorphism invariance), and emerges naturally as a Palatini formalism with the metric tensor and affine connection being independent fundamental (``co-ordinate'') fields. (For earlier work on gauge theories of gravitation see for example~\cite{utiyama56, kibble67, sciama62, hehl76, hayashi80}.)

The theory is shown to be inherently inconsistent unless a quadratic invariant in the momentum field --- conjugate to the affine connection --- is added to the covariant Hamiltonian  which generates a dynamic response of space-time to deformations relative to the de Sitter geometry. The strength of that invariant relative to the GR is adjustable by a dimensionless coupling (deformation~\cite{herranz07}) parameter $1/g_1$. That ``inertia'' of space-time generates a ``geometrical stress'' amending the dynamics of matter and of the geometry of space-time.

In this paper we briefly report on the possible cosmological impact of that theory. After a review of the canonical gauge theory of gravitation we present the modified Friedman equations pertinent to the quadratic-linear Lagrangian. A suitable normalization of the solution is ensures that present-day cosmological observables are reproduced with the standard~$\Lambda$ CDM parameter set.
We show that in its late epoch the universe exits into a dark energy era, and anticipate new types of early-epoch solutions incl. inflation and bouncing.

Finally we address how the de Sitter deformation adds a term  $\sim M_p^2/g_1$ to the cosmological constant that relieves the rigid relation between the cosmological constant and the vacuum energy density of matter encountered in GR. This provides a new aspect for resolving the huge discrepancy (of the order of $10^{120}$) between the theoretically predicted and the observed current value of the cosmological constant.

\section{The covariant canonical gauge theory of gravitation}
The generalized Einstein equation (Covariant Canonical Gauge Gravitation, CCGG equation) has been derived in~\cite{struckmeier17a} from the generic regular Hamiltonian density of matter in the Palatini formalism. This is the first order formalism with the metric and the affine connection being independent fundamental degrees of freedom. The resulting covariant Hamiltonian density,
\begin{equation} \label{CGGE}
 \tilde{\HCd}_{\mathrm{}} = \frac{1}{4g_1\, \sqrt{-g}}\,\tilde{q}\indices{_\eta^{\alpha\xi\beta}}\,\tilde{q}\indices{_\alpha^{\eta\tau\lambda}}\,g_{\xi\tau}\,g_{\beta\lambda} -
 g_2\,\tilde{q}\indices{_\eta^{\alpha\eta\beta}}\,g_{\alpha\beta}
+ \tilde{\HCd}_{\mathrm{matter}}
\end{equation}
is built from the kinetic portion of the free space-time field and the matter Hamiltonian $\tilde{\HCd}_{\mathrm{matter}}$, which is a scalar density, includes coupling to curved space-time. $\tilde{q}\indices{_\eta^{\alpha\xi\beta}} = q\indices{_\eta^{\alpha\xi\beta}} \, \sqrt{-g}$ is the canonical conjugate field (``momentum'') to the affine connection field.
(Our conventions are the signature $(+,\,-,\,-,\,-)$ of the metric, and natural units $\hbar = c = 1$.)
The set of basic assumptions of the theory is minimal:
\begin{itemize}
 \item Action principle
 \item Principle of General Relativity
 \item Regularity of the Lagrangian (i.e, the Hesse matrix of the Lagrangian is regular)
 \item Metric compatibility (i.e, the metric preserves lengths and angles)\footnote{This is \emph{not} a necessary requirement. It however significantly reduces the complexity of the theory.}.
\end{itemize}
Mathematically, the quadratic term in Eq.~\eref{CGGE} ensures regularity, i.e.\ the existence and reversibility of Legendre transformations between the Hamiltonian and Lagrangian pictures.  Physically, it endows space-time with kinetic energy and hence inertia that gives rise to ``geometrical stress''. The coupling constants $g_1$ and $g_2$
have the dimensions $[g_1] = 1$ and  $[g_2] = L^{-2}$.
\footnote{This choice~\cite{struckmeier17a} of the Hamiltonian combining the quadratic Riemann invariant and the linear Einstein-Hilbert term
ensures that the Schwarzschild metric is a solution of the system and hence compatibility with observations on galactic scale. A generic ansatz including all contractions of the Riemann tensor has been investigated in~\cite{benisty18a}.}

The canonical equations of motion~\cite{struckmeier18a} can be resolved for the momentum fields, yielding
\begin{equation} \label{def:momentum}
 q_{\eta\alpha\xi\beta} = g_1\,\left(R_{\eta\alpha\xi\beta} - \hat{R}_{\eta\alpha\xi\beta}\right)
\end{equation}
for the conjugate gauge field. Hereby
\begin{equation} \label{def:maxsymR}
\hat{R}_{\eta\alpha\xi\beta} = g_2 \left(g_{\eta\xi}\,g_{\alpha\beta} - g_{\eta\beta}\,g_{\alpha\xi} \right)
\end{equation}
is the Riemann tensor of the maximally symmetric 4-dimensional space-time with a constant Ricci curvature scalar $\hat{R} = 12g_2$.
It is also called the ``ground state geometry of space-time''~\cite{carroll99} which is the de Sitter (dS) or the anti-de Sitter (AdS) space-time for positive or negative sign of $g_2$, respectively. The momentum $q_{\eta\alpha\xi\beta}$ thus describes the deformation of space-time relative to the maximally symmetric, (A)dS, geometry.

By combining the canonical equations the consistency (CCGG) equation generalizing GR is derived~\cite{struckmeier17a, struckmeier18a}:
\begin{equation} \label{eq:modEinstein}
 g_1\left( R^{\alpha\beta\gamma\mu}\, R_{\alpha\beta\gamma\nu}
           - \quarter \delta^\mu_\nu \, R^{\alpha\beta\gamma\delta}\, R_{\alpha\beta\gamma\delta}\right)
           + \frac{1}{8\pi G}\, \left( R\indices{^\mu_\nu}
           - \onehalf \delta^\mu_\nu \, R - \delta^\mu_\nu \, \Lambda \right) = \theta\indices{^\mu_\nu}.
\end{equation}
$\theta\indices{^\mu_\nu}$ is the energy-momentum (stress) tensor of matter\footnote{Whether the stress tensor is the canonical or the metric one depends on the spin of the matter field and on the underlying internal symmetry group, see~\cite{struckmeier18a}.}. The equation includes the Einstein tensor and would boil down to the Einstein equation in the special case $g_1 = 0$. However, within the canonical transformation framework $g_1 = 0$ is forbidden, and, as we will see below, the limit $g_1 \rightarrow 0$ is not continuous!
In order to align with the syntax used in the Einstein's General Relativity, the coupling constants $g_1$ and $g_2$ are expressed in terms of the gravitational coupling constant $G$ and the cosmological constant $\Lambda$:
\begin{align}
   -g_1\,g_2 &\equiv \frac{1}{16\pi G} = \onehalf M_p^2 \label{def:constantsg1}\\
   6g_1\,g_2^2  &\equiv \frac{\Lambda}{8\pi G} \;\:= M_p^2\,\Lambda
   \label{def:constantsg3}
\end{align}
with the reduced Planck mass $M_p := 1/ \sqrt{8 \pi G} = 2.428 \times 10^{18}$~GeV.
Combining the above equations yields
\begin{equation} \label{def:constantsg32}
 \Lambda = -3g_2 = \frac{3M_p^2}{2g_1}
 ,
\end{equation}
i.e.\ we find the cosmological constant being generated by the (A)dS curvature. The parameter $g_1$ (or equivalently $g_2 = -M_p^2/2g_1$) is the deformation parameter of the geometry. In GR the cosmological constant is identified with the vacuum energy of the constituent particles~\cite{carroll99}, i.e.\ it is assumed that the vacuum expectation value (VEV) of Eq.~\eref{eq:modEinstein} is non-zero and homogeneous in 4D (A)dS geometry. In CCGG this leads to the effective cosmological term
\begin{equation} \label{def:Lambdaeff}
 \Lambda_{\mathrm{eff}} := \Lambda + 8\pi G\,\left(\theta_{vac}^{\mathrm{mat}}- \Theta_{vac}^{\mathrm{st}} \right) = \frac{3M_p^2}{2g_1} + 8\pi G\,\Theta_{vac}^{\mathrm{res}}.
\end{equation}
$\Theta_{vac}^{\mathrm{res}}$ denotes the residual value from two contributions, the vacuum energy densities of matter, $\theta_{vac}^{\mathrm{mat}}$, and space-time, $\Theta_{vac}^{\mathrm{st}}$.  Of course, since the overall VEV of the stress and strain tensors, $\Theta_{vac}^{\mathrm{res}}$, depends on the geometry of the underlying space-time, we do not expect $\Lambda_{\mathrm{eff}}$, unlike $g_2$, to be a constant.

The ``normal ordered'' version of the CCGG equation~\eref{eq:modEinstein} can then be recast into
\begin{equation} \label{eq:modEinsteineff}
 g_1\left( R^{\alpha\beta\gamma\mu}\, R_{\alpha\beta\gamma\nu}
           - \quarter \delta^\mu_\nu \, R^{\alpha\beta\gamma\delta}\, R_{\alpha\beta\gamma\delta}\right)
           + \frac{1}{8\pi G}\, \left( R\indices{^\mu_\nu}
           - \onehalf \delta^\mu_\nu \, R - \delta^\mu_\nu \, \Lambda_{\mathrm{eff}} \right) = \hat{\theta}\indices{^\mu_\nu},
\end{equation}
where the stress tensor $\hat{\theta}\indices{^\mu_\nu}$ vanishes in vacuum. In the following we will omit the caret ($\,\hat{ }\,$) on $\hat{\theta}$, and let $\Lambda$ denote the effective cosmological function $\Lambda_{\mathrm{eff}}$.

\section{The Friedman model}
A basic assumption of cosmology, the ``Cosmological Principle'', states that matter is homogeneously and isotropically distributed on cosmological scales, and the 4D space-time is then isotropic at any given point of the three-dimensional spatial slice normal to the temporal direction. This model universe~\cite{friedman22,weinberg72} is endowed with the Friedman-Lemaitre-Robertson-Walker (FLRW) metric with the line element (for flat 3D space)
\begin{equation}\label{def:RWmetric}
ds^2 = dt^2 - a^2(t) \left[dr^2 + r^2\left(d\theta^2 + \sin^2(\theta)\,d\varphi^2\right)\right].
\end{equation}
Hereby $t,r,\theta, \varphi$ are the \emph{global co-moving coordinates} mapping the entire universe. The parameter $a(t)$, characterizing the relative size of the spatial section of the metric as function of the cosmological time $t$, remains the only freedom left. If $t_0$ is the current age of the universe, $a(t_0) = 1$ applies to today. The material content of this model universe is a perfect fluid of classical particles and radiation. (Notice that spin and torsion effects that are inherent to CCGG are neglected here, and the connection is then Levi-Civita.) The corresponding energy-momen\-tum tensor for a perfect fluid made of classical matter with the density $\rho$ and pressure $p$ is symmetric:
\begin{equation} \label{def:perfectemtensor}
\theta = \sum_{i=r,m}\, \mathrm{diag}(\rho_i,-p_i,-p_i,-p_i).
\end{equation}
$\rho_i$ and $p_i$ are functions of the global time $t$ only, and $i$ tallys just two basic types of matter, namely particles (``dust'') and radiation. Of course, the particle matter is by itself a sum over all standard model particles and dark matter, but for the purpose of this study that level of detail is not required. Radiation, on the other hand, includes not only genuine photon energy density but also contribution from any kind of highly relativistic particles where mass is negligible compared to their kinetic energy.

For a perfect fluid the equation of state (EOS), i.e.\ the relation between the density $\rho_i$ and the pressure $p_i$, is assumed to have the generic form
\begin{equation} \label{EOSgeneric}
 p_i = \omega_i\, \rho_i,
\end{equation}
where $\omega_i \in \RB$ is a constant. For radiation (and neglecting interaction between these materials)~\cite{weinberg72}, $p_r = \onethird\rho_r$, hence for $i=r$ we have $\omega_r=\onethird$. With the definition $n_i \equiv 3(\omega_i+1)$ we get $n_r = 4$. For particle matter in equilibrium, $p_m = 0$, hence for $i=m$ we have $\omega_m=0$ and $n_m = 3$. This implies~\cite{reid02, weinberg72}:
\begin{equation} \label{scalinglaw}
 \rho_i (t) \sim a^{-n_i}  \qquad \Rightarrow \qquad \rho_i\,a^{n_i} = \mathrm{const}.
\end{equation}

By the symmetry of the Friedman universe,
the yet unspecified cosmological  function $\Lambda$ can only depend on the global time coordinate. The analysis is further simplified by the scaling ansatz\footnote{It can be shown that a constant $\Lambda$ is inconsistent with the quadratic traceless Riemann term in the CCGG equation because the latter is not covariantly conserved.}
\begin{equation} \label{def:f(a)}
\Lambda(a) =: \lambda_0 \, f(a),
\end{equation}
with a dimensionless function $f(a)$ and some constant $\lambda_0$ to be specified later.
It describes deviations from the scaling behavior \eref{scalinglaw} and interactions between the components of matter.

It is convenient now to define the quantities
\begin{equation} \label{scalinglaw1}
 C_i := \frac{8\pi G}{3}\,\rho_i\,a^{n_i} = \mathrm{const.}, \qquad i=r, m.
\end{equation}
and
\begin{align}
   C_\Lambda &:= \onethird\lambda_0\  \qquad \omega_\Lambda = -1, \,n_\Lambda = 0. \label{def:CLambda}
\end{align}
The standard Friedman equation as known from General Relativity,
\begin{equation} \label{eq:firstFriedman}
  H^2(a) = \frac{8\pi G}{3}  \sum_{i=r,m}\rho_i + \onethird\lambda_0 = \frac{8\pi G}{3}  \sum_{i=r,m,\Lambda}\rho_i,
\end{equation}
is modified in CCGG. After some straightforward algebra\footnote{The details of the derivation are omitted here in this short version but given in the full paper (Vasak et al. in preparation).} we obtain:
\begin{equation}
  H^2 = \sum_{i=r,m}\,C_i\,a^{-n_i} + C_\Lambda\,f(a)
  +  g_1\,\frac{32\pi G M(a)}
   {\left(1-32\pi G g_1 M(a)\right)}\, \sum_{i=r,m}\,\quarter C_i\,n_i\,a^{-n_i}. \label{eq:modFriedman3}
\end{equation}
The first term on the r.h.s. is the contribution from conventional matter, i.e.\ dust and radiation, while the second term stands for the non-standard scaling of matter expressed by the cosmological function~$C_\Lambda \, f(a)$. The third, geometry, term can be traced back to the quadratic momentum invariant in the Hamiltonian. The term
\begin{equation}
   M(a) = \quarter C_m\,a^{-3} + C_\Lambda\,f(a), \label{def:M1}
\end{equation}
is essentially the Ricci tensor of the FLRW metric. Using the relation Eq.~\eref{def:constantsg1} (and suppressing the dependence of $M$ and $f$ on~$a$) this becomes:
 \begin{align} \label{eq:modFriedman4}
 H^2 &= \sum_{i=r,m,K}\,C_i\,a^{-n_i} + C_\Lambda\,f + \frac{M}{\onehalf g_2-M}\,(\threequarter C_m\,a^{-3}+ C_r\,a^{-4}) \\
 &= \sum_{i=r,m,K}\,C_i\,a^{-n_i} + C_\Lambda\,\chi(a) \nonumber.
\end{align}
We call the function
\begin{equation} \label{def:h(a)2}
 \chi(a) := f(a) +
 \frac{
 \left( C_\Lambda \,f(a) + \quarter a^{-3}\,C_m \right) \left( \threequarter a^{-3}\,C_m + a^{-4}\,C_r \right)
 }
 {C_\Lambda \, \left(
 \onehalf g_2 - \quarter a^{-3}\,C_m -C_\Lambda \,f(a)
 \right)}.
 \end{equation}
 ``dark energy function''. It absorbs all non-Einsteinian contributions into a novel ``geometrical stress'' term. Hence we call 
 \begin{equation} \label{def:rhoDE}
  \rho_{DE} =: \frac{3C_\Lambda}{8\pi G}\,\chi(a)
\end{equation}
``dark energy density''.
 It is well defined since the function $f(a)$ obeys the unique differential equation
\begin{equation} \label{ODE:f(a)}
 \frac{df}{da} =
  \frac{3C_m}{4C_\Lambda} a^{-4} \, \times
  \frac{
  \onehalf g_2 \left(\threequarter C_m a^{-3} + C_r a^{-4}\right)
  - \left(\onehalf g_2 - \threequarter C_m a^{-3} - C_\Lambda \,f \right)
  \left(\threequarter C_m a^{-3} + C_\Lambda \,f \right)
    }
  {\onehalf g_2 \left(\threequarter C_m a^{-3} + C_r a^{-4}\right)
    +   \left(\onehalf g_2 - \threequarter C_m a^{-3} - C_\Lambda \,f \right)^2
  }.
\end{equation}

Aligning the dark energy function with the observed present-day value of the cosmological constant requires setting $\lambda_0 = \Lambda_{\mathrm{obs}}$ and $\chi_0 \equiv \chi(a=1) = 1$.  For $f_0 \equiv f(1)$ this implies
\begin{equation} \label{eq:h0}
 \chi_0 = f_0 +
 \frac{
 \left( C_\Lambda \,f_0 + \quarter C_m \right) \left( \threequarter C_m + C_r \right)
 }
 {C_\Lambda \, \left(
 \onehalf g_2 - \quarter C_m -C_\Lambda \,f_0\right)
 } = 1.
 \end{equation}
Equation~\eref{eq:h0} is a quadratic equation for $f_0$ with two roots for a given $g_2$ but yields a unique expression for $g_2$ in terms of $f_0$:
 \begin{equation} \label{prooffne1}
  \onehalf \, g_2(f_0) = \frac{C_\Lambda}{f_0-1} \, \left[
  f_0^2 - \frac{f_0}{C_\Lambda} (\onehalf\,C_m+C_r+C_\Lambda)
  - \frac{C_m}{ 4C_\Lambda^2 } (\threequarter \,C_m +C_r+C_\Lambda)\right].
 \end{equation}
With Eq.~\eref{def:constantsg1} an equivalent expression for $g_1$ is obtained:
 \begin{equation} \label{prooffne2}
  g_1(f_0) = \frac{f_0 - 1}{32 \pi G C_\Lambda} \, \left[
  f_0^2 - \frac{f_0}{C_\Lambda}   (\onehalf\,  C_m + C_r + C_\Lambda)
  - \frac{C_m}{4C_\Lambda^2} (\threequarter\, C_m + C_r + C_\Lambda) \right]^{-1}.
 \end{equation}
 \section{The dark energy function in the CCGG-Friedman model}
With the equations \eqref{def:rhoDE} and \eqref{eq:h0}, the dark energy density
\begin{equation} \label{def:rholambda1}
 \rho_{DE}(a) = 
 \frac{3C_\Lambda}{8\pi G}\,\chi(a) = \frac{\lambda_0\,f(a)}{8\pi G} + \frac{3}{8\pi G}
 \frac{M}{\onehalf g_2-M}\,\left( \threequarter C_m\,a^{-3}+ C_r\,a^{-4} \right)
\end{equation}
just generalizes the conventional notion of dark energy, substituting the cosmological constant by contributions from non-standard matter and from space-time geometry. For dark energy pressure we obtain analogously
\begin{equation} \label{def:plambda1}
 p_{DE}(a) =  -\frac{\lambda_0\,f(a)}{8\pi G} + \frac{1}{8\pi G}
 \frac{M}{\onehalf g_2-M}\,\left( \threequarter C_m\,a^{-3}+ C_r\,a^{-4} \right).
\end{equation}
Then the EOS is
\begin{equation} \label{def:omegaDE}
 \omega_{DE} (a) = \frac{p_{DE}(a)}{\rho_{DE}(a)} = - \frac
 {\lambda_0 f(a) - \frac{M}{\onehalf g_2-M}\,\left( \threequarter C_m\,a^{-3}+ C_r\,a^{-4} \right)}
 {\lambda_0 f(a) + \frac{3M}{\onehalf g_2-M}\,\left( \threequarter C_m\,a^{-3}+ C_r\,a^{-4} \right)}.
\end{equation}
The total density including all components of matter and dark energy, \eqref{eq:firstFriedman}, and the corresponding pressure and EOS read accordingly
\begin{align} 
 \rho_{tot}(a) &= \frac{8\pi G}{3} \sum_{i=r,m,DE}\rho_i = \rho_{DE}(a) + \frac{3}{8\pi G} \, \left(C_m\,a^{-3} + C_r\,a^{-4} \right) \label{def:rhotot1}\\
    p_{tot}(a) &= \frac{8\pi G}{3} \sum_{i=r,m,DE}p_i = p_{DE}(a) +  \frac{1}{8\pi G} \, C_r\,a^{-4} \label{def:ptot1} \\
\omega_{tot} (a) &= \frac{p_{tot}(a)}{\rho_{tot}(a)} = 
\frac{p_{DE}(a) +  \frac{1}{8\pi G} \, C_r\,a^{-4}}{\rho_{DE}(a) + \frac{3}{8\pi G} \, \left(C_m\,a^{-3} + C_r\,a^{-4} \right)}.\label{def:omegatot}
 \end{align}

\section{The $\Lambda$ CDM parameters and the quadratic Riemann theory} \label{sec:lambdaCDM}
As is common in conventional cosmology we re-define the constants $C_i$ in Eqs.~\eref{scalinglaw1} and~\eref{def:CLambda} in relation to the critical density expressed by the Hubble constant $H_0 := H(1)$, to
\begin{equation} \label{def:omegai}
 \Omega_i := \frac{C_i}{H_0^2}.
\end{equation}
Then the Friedman equation can, for $a = 1$, be recast into
 \begin{equation} \label{eq:secondFriedman2}
 1 = \sum_{i=r,m,\Lambda}\,\Omega_i,
\end{equation}

The construction~\eref{def:h(a)2} of the function $\chi(a)$ in the CCGG-Friedman equation  ensures that for the CCGG-Friedman model the equation
\begin{equation} \label{eq:modFriedman42}
 1 = \sum_{i=r,m}\,\Omega_i + \Omega_\Lambda\,\chi(1)
\end{equation}
holds. With $\chi(1) = 1$, Eq.~\eref{eq:modFriedman42} formally coincides with the conventional Einstein-Friedman equation~\eref{eq:secondFriedman2}. Hence we can adopt the parameters of the conventional Einstein-Friedman model based on the standard model of elementary particles, cold dark matter, and a constant cosmological constant~$\Lambda_{\mathrm{obs}}$. For the present-day energy/mass densities of matter including dark matter, of radiation including photons and neutrinos, and for the cosmological constant we have (see for example~\cite{planck15,bernal17}):
 \begin{align*}
C_m &= \frac{8\pi G}{3}\,\rho_m \approx 6.790 \times 10^{-85}\, \mathrm{GeV^2}\\
C_r &= \frac{8\pi G}{3}\,\left(\rho_\gamma + \rho_\nu\right) \approx 1.898 \times 10^{-88}\, \mathrm{GeV^2}\\
C_\Lambda &= \onethird\,\Lambda_{\mathrm{obs}} \approx 1.567 \times 10^{-84}\, \mathrm{GeV^2}.
\end{align*}
From Eq.~\eref{eq:modFriedman4} we get for $a = 1$ the standard value of the Hubble constant $H_0 \equiv H(a=1)$:
\begin{equation}
  H_0^2 = \sum_{i=r,m,K}\,C_i + C_\Lambda\ \approx 1.5 \times 10^{-42} \mathrm{GeV} \sim 70 \mathrm{\frac{km}{s \,Mpc}},
  \label{eq:modFriedman43}.
\end{equation}
often expressed as the $h_0$ relative to $H_{100}$:
 \begin{align}
 H_0 &= h_0 \, H_{100} \\
 H_{100} &= 100 \mathrm{\frac{km}{s \, Mpc}} = 2.135 \times 10^{-42}\,\mathrm{GeV}. \nonumber
 \end{align}
\begin{figure}[ht]
\includegraphics[width=\linewidth]{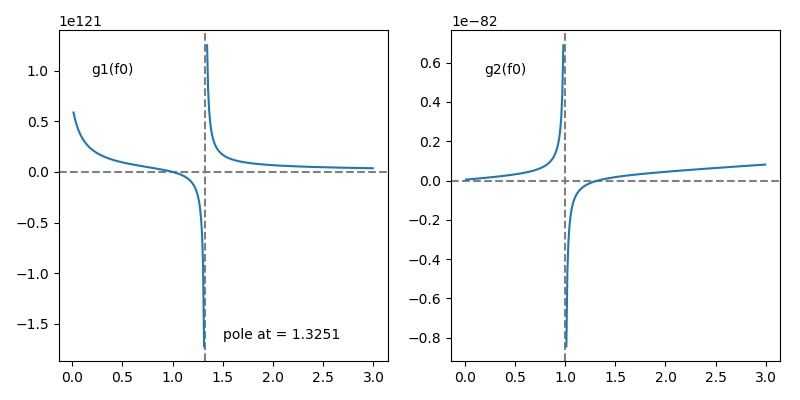}
\caption{
 The functions $g_1(f_0)$ (left) and $g_2(f_0)$ (right) for positive $f_0$ are displayed. Notice the different scales, $10^{121}$ for $g_1$ and $10^{-82}$ for $g_2$.
 The function $f_0(g_2)$ has two roots, hence $g_2(f_0)$ has two branches. For $f_0 \rightarrow 1_-$ ($1_\pm$ denotes $1\pm\epsilon$ with $\epsilon \ll 1$.) the constant $g_1$ will be positive and close to zero, while for $f_0 \rightarrow 1_+$ we find a negative and small $g_1$. Then $g_2$ approaches $\pm \infty$, respectively. A large and positive $g_1$ requires  $f_0 \ge (3C_m/4+C_r+C_\Lambda)/C_\Lambda \approx 1.325$. For $f_0 < (3C_m/4+C_r+C_\Lambda)/C_\Lambda$ we find a negative coupling constant $g_1$ and hence also a negative $g_2$.
 In the plot we do not show the branch for negative $f_0$ where $g_1 < 0$.}
 \label{fig:g1(f0)}
\end{figure}
An important cosmological ``observable'' is also the dimensionless, so called deceleration function
\begin{equation} \label{eq:deceleration}
q =: -\frac{\ddot{a}}{\dot{a}^2}\,a \equiv -\frac{\ddot{a}}{a}\,\frac{1}{H^2} = -\frac{2M}{H^2} +1,
\end{equation}
which implicitly depends on the quadratic term. Written out explicitly, its present-day value~$q_0 := q(1)$ reads
\begin{equation}\label{eq:decelerationf0}
q_0 = 1 -\onehalf\Omega_m - \Omega_K - \twothird \lambda_0 f_0.
\end{equation}
In order to align the theoretical value of $q_0$ with the observed value, $q_0 = -0.5$, $f_0$ must be close to~$1$. (How close, depends on the accuracy of the value of $q_0$ derived from observations that is still under discussion~\cite{planck15, bernal17}.) Hence the parameter $f_0$ determines via Eq.~\eref{prooffne2} the coupling constant $g_1$ --- and hence deformation parameter $g_2 = M_p^2/2g_1$, c.f.\ \fref{fig:g1(f0)}.  Obviously, observations can be used to infer the value of the key parameter of the theory, $g_1$.  It is important to realize that $f_0 = 1$ is possible if and only if $g_1 = 0$, i.e.\ only in the realm of the standard Einstein-Friedman model, and the limiting process, $g_1 \rightarrow 0$, is not continuous!

Thus, within the present accuracy of observations, the CCGG-Friedman model can be made in this approximation consistent with the
``$6$-parameter'' or ``Concordance'' model~\cite{planck15} (``$\Lambda$ Cold Dark Matter'', $\Lambda$CDM) of the standard Einstein-Friedman cosmology,
\begin{equation}
\left(\Omega_m, \, \Omega_r, \, \Omega_\Lambda, \, \Omega_K, \,h_0, \,q_0\right)_{\mathrm{standard}} \approx
\left(0.3, \,0, \,0.7, \,0, \,0.70, \,-0.5\right).
 \end{equation}
This set represents a flat (open or closed) universe currently undergoing an accelerated expansion. (We have set~$K \approx 0$ throughout this paper anticipating consistency with the observed CBM radiation isotropy~\cite{planck15}.)
\section{Dark energy and inflation}
In the limit $a \rightarrow \infty$ Eq.~\eref{ODE:f(a)} shows $f(a) \rightarrow \mathrm{const.}$
Then, by the definition~\eqref{def:h(a)2} of $\chi$, we find $\chi \rightarrow f$ and recover in the late epoch the dark energy scenario with a modified cosmological constant. However, as the second term in~\eref{def:h(a)2} may become dominant or even singular if the denominator approaches zero, the early dynamics of the universe will be fundamentally different from GR. First analyses show that this is indeed the case~\cite{benisty18b}, and that bouncing and inflation are possible features of the theory.

\medskip
To shed more light on the solution landscape we consider here the following limiting cases:
\begin{enumerate}
 \item Because of the relation (\ref{def:constantsg1}), $g_1 \rightarrow 0_\pm$  is equivalent to $g_2 \rightarrow \pm \infty$. In case that $f(a)$ remains finite the second term in \eref{def:h(a)2} would vanish and $\chi = f \equiv 1$ must hold. 
 Dark energy is then replaced by the cosmological constant $\lambda_0$ with the EOS $\omega_{DE} = -1$, as in the conventional Friedman model. In the early epoch, though, singularities in the function $\chi$ are shown to exist in first numerical calculations. 
 \item On the other hand, with positive $f_0$ and $g_1 \rightarrow \pm \infty$, the (A)dS Ricci scalar vanishes as $g_2 \rightarrow \mp 0$, and \eref{ODE:f(a)} becomes
 \begin{equation}
  f' = \threequarter \frac{C_m}{C_\Lambda}\, a^{-4},
 \end{equation}
which, with the initial condition $f(1) = f_0$, can readily be integrated to
\begin{equation}
 f(a) = \frac{C_m}{4C_\Lambda} (1-a^{-3}) + f_0.
\end{equation}
This implies, via \eref{def:h(a)2},
\begin{equation} \label{eq:h(a)forg_2=0}
 \chi(a) = f_0 + \frac{1}{C_\Lambda}\left( \quarter C_m - C_m\,a^{-3} - C_r\,a^{-4} \right).
\end{equation}
From the boundary condition, $\chi(1) = 1$, we conclude
\begin{equation} \label{eq:f_0forg_2=0}
 f_0 = \frac{\threequarter C_m + C_r + C_\Lambda}{C_\Lambda},
\end{equation}
and hence
\begin{equation}
 C_\Lambda\,\chi(a) = C_m + C_r + C_\Lambda - C_m\,a^{-3} - C_r\,a^{-4}.
\end{equation}
The dark energy function leads asymptotically to a universe dominated by the cosmological constant 
\begin{equation}
\onethird \Lambda_{\mathrm{\infty}}   = C_m + C_r + C_\Lambda.
 \end{equation}
The total energy density and pressure, with the contributions from matter and radiation added, become constant, and
the resulting EOS, $\omega_{tot}$, approaches -1 for large $a$:
\begin{align}
 \rho_{tot} &\rightarrow \qquad \frac{3}{8 \pi G}\,(C_m + C_r + C_\Lambda), 
 \nonumber \\
                                  p_{tot} &\rightarrow \quad -\frac{3}{8 \pi G}\,(C_m + C_r + C_\Lambda) 
							    \nonumber \\
                             \omega_{tot} &\rightarrow \quad -1. 
							    \nonumber					    
\end{align}
In this case, the late universe is dark energy dominated with a cosmological constant given by the total density of all constituents, i.e. matter and radiation have been transformed into dark energy. Again, singularities of $\chi(a)$ in the early epoch cannot be excluded at this point.
 \item In a universe without particle matter, $C_m = 0$, the Friedman equation can be integrated to
  \begin{equation} \label{def:h(a)nom2}
 \chi(a) = f_0 +  (1-f_0) \,a^{-4}.
 \end{equation}
The energy density and pressure of the dark energy is then
\begin{align}
 \rho_{DE} &= \frac{3 C_\Lambda\,f_0}{8 \pi G}\,
 \left( 1 + \frac{C_r\,a^{-4}}{\onehalf g_2-C_\Lambda f_0} \right) = \frac{3 C_\Lambda}{8 \pi G}\,\left[f_0 +(1-f_0)a^{-4} \right]
 \nonumber \\
                   p_{DE} &= -\frac{3C_\Lambda\,f_0}{8 \pi G}\,
 \left( 1 - \frac{\onethird C_r\,a^{-4}}{\onehalf g_2-C_\Lambda f_0} \right)
 =-\frac{3 C_\Lambda}{8 \pi G}\,\left[f_0 -\onethird (1-f_0)a^{-4} \right].
 \nonumber
\end{align}
For $a \rightarrow 0$ dark energy is transformed into radiation, and $\omega_{DE} \rightarrow \onethird$, while for $a \rightarrow \infty$ we recover the EOS of standard dark energy,~$\omega_{DE} \rightarrow -1$.
\item Setting $\lambda_0 = 0$ we are left with the pure geometry term in the dark energy function. As long as $C_m \ne 0$ the energy density and pressure of such a ``geometrical fluid'', 
\begin{align}
 \rho_{DE}(a) &= \frac{3}{8\pi G} \,\frac{\quarter C_m\,a^{-3}}{\onehalf g_2-\quarter C_m\,a^{-3}}\,
 \left( \threequarter C_m \,a^{-3} + C_r\,a^{-4} \right) \nonumber \\
                         p_{DE}(a) &= \frac{1}{8\pi G} \,\frac{\quarter C_m\,a^{-3}}{\onehalf g_2-\quarter C_m\,a^{-3}}\,
 \left( \threequarter C_m \,a^{-3} + C_r\,a^{-4} \right) \nonumber 
 \end{align}
yield a radiation-like EOS, $\omega_{DE} = \onethird$. However, then the total energy density can become negative violating all energy conditions, and dark energy vanishes asymptotically for $a \rightarrow \infty$. We conclude that such a configuration is at odds with observations, and $\lambda_0 \ne 0$ is necessary.
\item In the dark energy dominated case, $C_m = C_r = 0$ and $\lambda_0 = \Lambda_{\mathrm{obs}}$, $f'(0) = 0$ and the dark energy energy function reduces to the cosmological constant. Obviously, the presence of matter is necessary for the dark energy that emerges from the space-time's inertia to exist.  
\end{enumerate}

\section{The cosmological constant problem}
The quadratic ``kinetic'' term in the Hamiltonian and Lagrangian densities adds the (A)dS curvature parameter $g_2 = M_p^2/2 g_1$ to the cosmological constant.
The cosmological constant is then composed of two \emph{independent} portions: The vacuum energies of space-time and matter, and  the ``(A)dS curvature term''. The latter provides an additional freedom to align the theoretical and observational values of the cosmological constant, and resolves the so called ``Cosmological Constant Problem''. Any present-day vacuum energy density $\Theta_{vac}^{\mathrm{res}}$ can, by a ``suitable'' choice of the \emph{necessarily} non-vanishing (A)dS deformation parameter, be made compatible with the present-day value of the cosmological constant.
\section{Conclusion and outlook}
 The CCGG theory extends General Relativity and facilitates, in a mathematically rigorous way, a consistent description of the dynamics of space-time and of the evolution of the universe from start to end. The theory unambiguously fixes the coupling of space-time to matter fields~\cite{struckmeier17a} and naturally introduces a new fundamental deformation parameter proportional to the Planck mass.  Such a construction of space-time~\cite{wise10} has been discussed earlier under the heading of de Sitter relativity to calculate the cosmological constant, and explain cosmic coincidence and time delays of extragalactic gamma-ray flares (see for example~\cite{aldrovandi09}). On the other hand, the Schwarzschild metric is also a solution of the CCGG equations such that the standard tests on the scale of the solar system are met \cite{struckmeier17a}.

 This paper is just a brief summary of the current status of investigating the CCGG cosmology. Further  studies are underway~\cite{benisty18a, benisty18b, vasak18b}
 using the Friedman model in the spirit of modified gravity models (cf.\ for example~\cite{starobinsky80,tsujikawa10,capozziello03}). The objective is to analyze the geometrically induced portion of Dark Energy (``geometrical fluid''), and to identify observables to detect the presence and measure the relative contribution of the quadratic term.
 However, since the deformation parameter enters also other observables, a more comprehensive study is required to identify a parameter set that is consistent with observations.
\section*{References}
\input{Group32_02.bbl}

\end{document}

%% file: Group32_02.bbl
\providecommand{\newblock}{}